\documentclass[authoryear,12pt]{elsarticle}
\usepackage{graphicx}
\usepackage{float}
\def\lsim{\ ^{<}\!\!\!\!_{\sim}\>}
\def\gsim{\ ^{>}\!\!\!\!_{\sim}\>}

\begin{document}


\centerline{\large\bf Dynamical evolution and end states}

\centerline{\large\bf of active and inactive Centaurs}

\vspace{2cm}

\centerline{Julio A. Fern\'andez$^{(1,*)}$, Michel Helal$^{(1)}$, Tabar\'e Gallardo$^{(1)}$}

\bigskip

\noindent (1) Departamento de Astronom\'ia, Facultad de Ciencias, Universidad de la Rep\'ublica, Igu\'a 4225, 14000 Montevideo, Uruguay

\vspace{2cm}


\noindent Number of pages: 24\\
Number of figures: 14\\
Number of tables: 3

\vspace{2cm}


\vspace{2cm}

\centerline{Planetary and Space Science, in press}

\vspace{4cm}

\noindent $^{(*)}$ Corresponding author\\
email: julio@fisica.edu.uy

\vfill
\eject

\centerline{\bf Abstract}

\bigskip

We numerically study the dynamical evolution of observed samples of active and inactive Centaurs and clones that reach the Jupiter-Saturn region. Our aim is to compare the evolution between active and inactive Centaurs, their end states and their transfer to Jupiter family comets and Halley-type comets. We find that the median lifetime of inactive Centaurs is about twice longer than that for active Centaurs, suggesting that activity is related to the residence time in the region. This view is strengthened by the observation that high-inclination and retrograde Centaurs (Tisserand parameters with respect to Jupiter $T_J < 2$) which have the longest median dynamical lifetime ($=1.37 \times 10^6$ yr) are all inactive. We also find that the perihelion distances of some active, comet-like Centaurs have experienced drastic drops of a few au in the recent past ($\sim 10^2-10^3$ yr), while such drops are not found among inactive Centaurs. Inactive Centaurs with $T_J \lsim 2.5$ usually evolve to Halley-type comets, whereas inactive Centaurs with $T_J \gsim 2.5$ and active Centaurs (that also have $T_J \gsim 2.5$) evolve almost always to Jupiter family comets and very seldom to Halley type comets. Inactive Centaurs are also more prone to end up as sungrazers, and both inactive and active Centaurs transit through different mean motion resonances (generally with Jupiter) during their evolution.

\vspace{1cm}


\vfill
\eject

\section{Introduction}

Centaurs are objects moving between the orbits of Jupiter and Neptune on unstable orbits. They are assumed to be a transient population between the trans-neptunian population and that reaching the inner planetary region \citep{Levi97,Tisc03}. Conventionally, Centaurs are defined as those objects having perihelion distances in the range $5.2 < q < 30$ au, and semimajor axes $a < 30$ au, leaving aside Jupiter's Trojans (bodies in the 1:1 Mean Motion Resonance (MMR) with Jupiter) \citep{Jewi09}. Their mean dynamical lifetime in the range $5.2 < q < 30$ au is found to be 72 Myr \citep{Disi07}. Most Centaurs are escapees from the Scattered Disk \citep{Dunc97,Disi07}, though a minor fraction may come from Plutinos decoupled from the 2:3 MMR with Neptune \citep{Disi10}. Once Centaurs approach Saturn their dynamical evolution speeds up. For a sample of Centaurs skewed to objects with perihelia in the Jupiter-Uranus zone, \citet{Tisc03} obtained a median dynamical lifetime of 9 Myr.

Once a large enough sample of Centaurs was assembled, a bimodality in their colors became apparent \citep{Peix03}, being some of them reddish while others showed more neutral spectra. From WISE observations of a sample of 52 Centaurs, \citet{Baue13} derived a mean albedo of $0.08 \pm 0.04$, namely they are in general dark objects like comets. Their colors showed a bimodal distribution, confirming previous results. The majority showed neutral, solar-like colors with $B-R < 1.4$ mag, while a minor fraction showed redder colors clustering around $B-R \sim 1.85$ \citep{Baue13}. From Herschel infrared observations of trans-neptunian objects and Centaurs \citet{Lace14} find a strong correlation between color and albedo that goes from bright red objects with mean geometric (visual) albedo $p_v \sim 0.15$ to dark neutral ('gray') objects with a mean $p_v \sim 0.05$.

What is striking is that some Centaurs look inactive (asteroidal) so they have received asteroid designations, while others show activity, in some cases so conspicuous that they have been designed with comet names since their discovery. There are also a few objects that show intermittent or sporadic activity, so they have received dual comet-asteroid designations. Centaur 2060 Chiron (with the dual comet designation 95P) is an emblematic case of a body first observed inactive, and thus classified as an asteroid, but that later showed a coma apparently produced by one of two regions rich in CO \citep{Meec90}. Another good example is 60558 Echeclus (2000 EC98) first classified as an inactive Centaur until it experienced a massive cometary outburst at a heliocentric distance $r \sim 13$ au with an increase of $\sim 5$ magnitudes \citep{Choi06}.

It is believed that highly volatile gaseous species, like CO or CO$_2$, drive the activity in active Centaurs, since at the heliocentric distances where these bodies move it is too cold for water ice to sublimate. Indeed, CO was observed at submillimeter wavelengths in 29P/Schwassmann-Wachmann 1 by \citet{Sena94} and more recently in 174P/Echeclus at 6 au \citep{Wier17}. Several other searches of CO outgassing were carried out in several other Centaurs, including the weakly active 2060 Chiron and 166P/NEAT, leading to negative results which allowed the authors to set upper limits to the CO gas production rate \citep{Bock01,Jewi08,Drah17}. These upper limits (equivalent to $\sim 0.1-1$\% or less of the surface area occupied by CO ice) suggest that the activity is too low to be due to the sublimation of CO ice exposed on the surface of the objects, so other mechanism, like the release of CO during the phase transition from amorphous to crystalline ice, may be at work \citep{Jewi08}. We will turn back to this point in the Discussion. 

The colors of Centaurs appear to be related to the presence -or not- of ultrared matter on their surfaces, namely complex organic molecules irradiated by cosmic rays and solar wind particles. A competing process is surface erosion by micrometeorite impacts or outgassing able to excavate and blanket the surface with neutral ('gray') dust particles. \citet{Jewi15} finds that the disappearance of the ultrared matted starts at about 10 au and that this phenomenon is also related to the onset of activity in some Centaurs. \citet{Meli12} find that the observed color bimodality of Centaurs is likely related to the observed thermal processing of their surfaces, where the red objects show a tendency to penetrate less in the inner regions of the solar system. So color and activity may be correlated, but the story is not so simple: while objects of the red group are inactive, we find active as well as inactive Centaurs within the gray group \citep{Jewi09,Jewi15}. This observation compels us to explore the causes of activity independent of color.   

We plan to study the dynamical evolution of the observed objects with perihelion distances $4 < q < 9$ au and aphelion distances $Q < 30$ au, namely those objects that reach the region around Jupiter's orbit, or the Jupiter-Saturn region and move entirely inside Neptune's orbit. The selected sample does not match the usual definition of Centaur, as formulated before, since we included objects crossing Jupiter's orbit down to $q = 4$ au. Nevertheless we think that our selected sample does not differ in nature from the conventional Centaurs since most of them cross Jupiter's orbit back and forth many times during their orbital evolution. A paradigmatic case is the comet/Centaur 39P/Oterma that switched from a quasi-Hilda orbit with $q=3.4$ au, lying entirely inside Jupiter's orbit, to a Centaur-like orbit with $q=5.5$ au after a close encounter with Jupiter in 1963.

This paper can be seen as a continuation of a previous work \citep{Fern16} where the authors showed that long-period comets (LPCs) are not a suitable source of Halley-type comets (HTCs), and that the latter may mainly come from another source that was identified with Centaurs. The difference is that in this case we start the evolution with Centaurs that already reached the Jupiter-Saturn region and focus on the differences in the dynamical evolution between active and inactive Centaurs. Our aim is twofold: 1) to analyze their end states and their role as precursors of HTCs and Jupiter family comets (JFCs), and 2) to find out if there are significant differences in the evolution of active and inactive Centaurs that may help to explain the observed activity in some of these bodies.

\section{The observed samples}

We extracted 68 inactive and 55 active Centaurs from NASA's Solar System Dynamics $(https://ssd.jpl.nasa.gov/sbdb\_query.cgi)$ with perihelion distances in the range $4 < q < 9$ au, aphelion distances $Q < 30$ au, and quality code $\leq 5$ (in a scale going from 0 : very good to 9 : very poor). Jupiter's Trojans have been removed from our samples. On the other hand, as mentioned above, we included Jupiter-crossing objects that conventionally do not enter into the category of Centaur. The advantage of limiting our sample to Jupiter approachers or crossers is that they evolve faster, on a time scale of about one Myr, so we can carry out numerical simulations with massive samples on a reasonable computer time scale of about one month. 

\begin{figure}[h]
\resizebox{12cm}{!}{\includegraphics{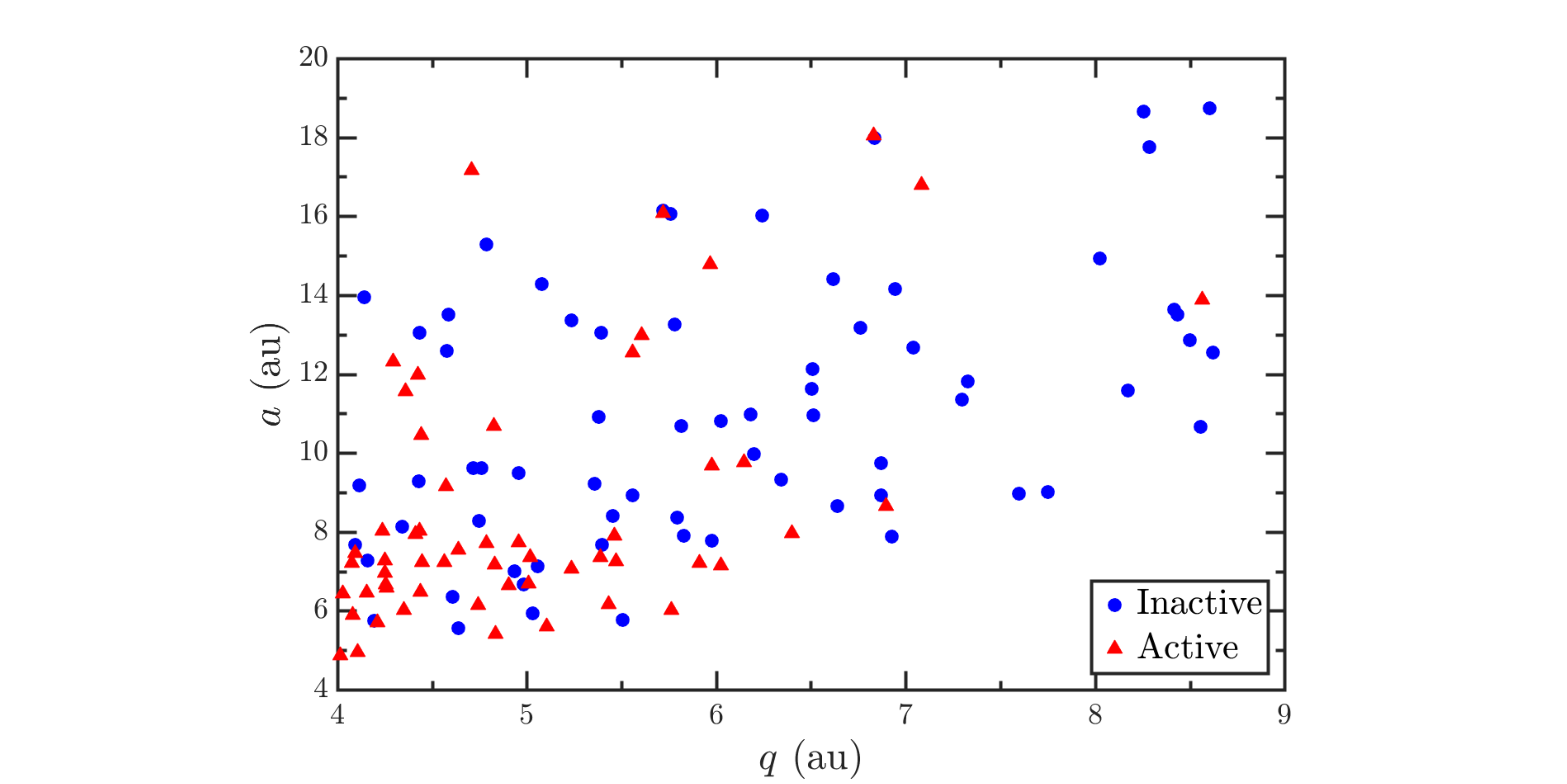}}
\caption{Plot of the perihelion distances $q$ versus semimajor axes $a$ for the observed samples of inactive and active Centaurs described in Section 2.}
\label{Centaurs_aq}
\end{figure}

\begin{figure}[h]
\resizebox{12cm}{!}{\includegraphics{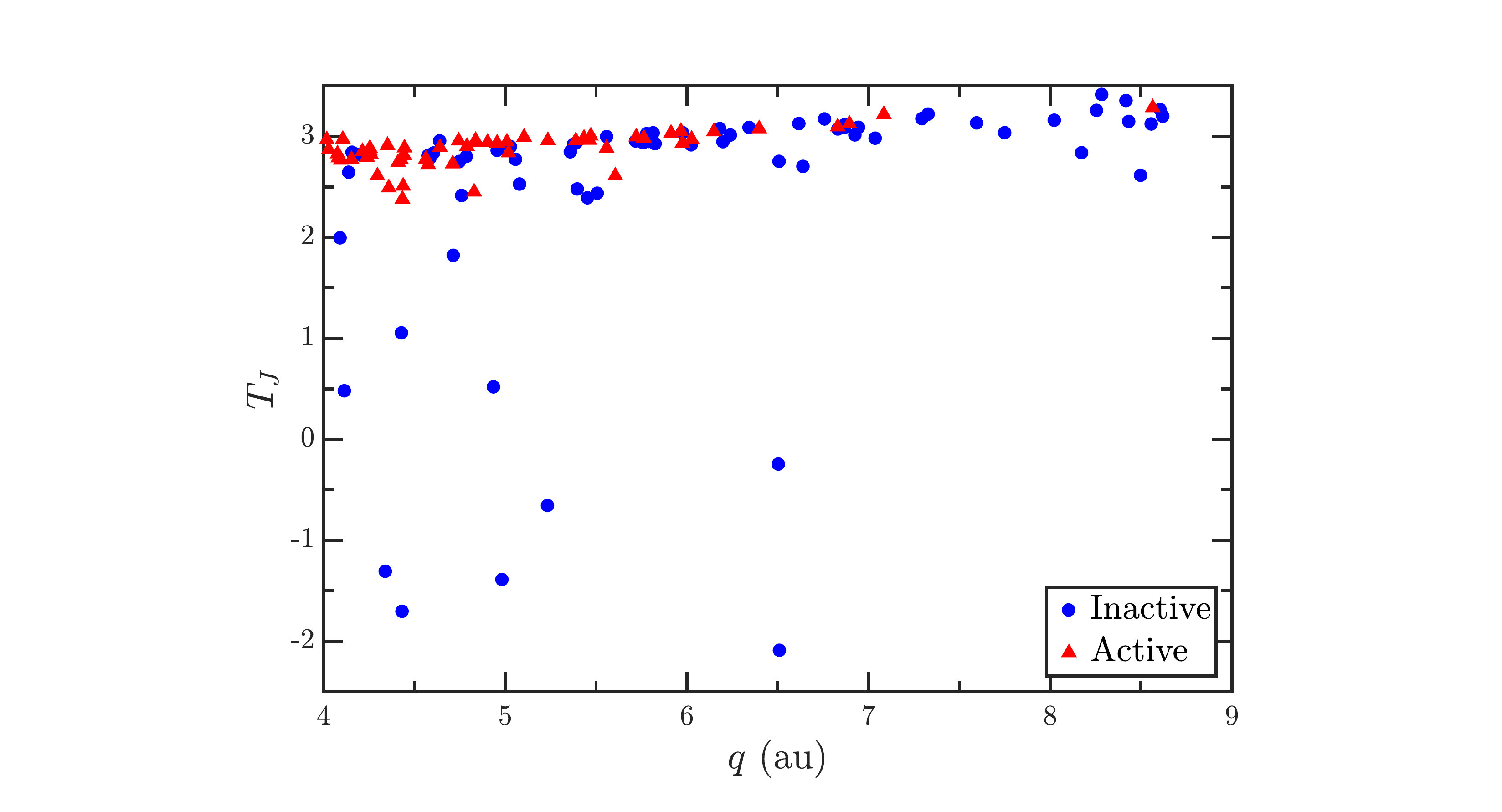}}
\caption{Plot of the perihelion distances $q$ versus the Tisserand parameters with respect to Jupiter $T_J$ for the observed samples of inactive and active Centaurs described in Section 2.}
\label{Centaurs_iq}
\end{figure}

We show in Fig. \ref{Centaurs_aq} the distribution of the observed inactive and active Centaurs in the plane perihelion distance versus semimajor axis $a$. Both populations are rather well mixed, though we notice some segregation of active Centaurs toward small $q$ values, while the mean perihelion distance of inactive Centaurs is $<q> = 6.05$ au, the corresponding value for active Centaurs is $<q> = 5.03$ au. Fig. \ref{Centaurs_iq} shows a plot of the Tisserand parameter with respect to Jupiter ($T_J$) versus perihelion distances $q$ for the active and inactive Centaurs. We see that most Centaurs are confined within the strip $2.5 \lsim T_J \lsim 3.5$ for different $q$. These Centaurs also have low inclinations $i \lsim 30^{\circ}$. On the other hand, Centaurs with $T_J \lsim 2$ move in high-inclination or retrograde orbits. For objects that do not closely approach Jupiter (say $q \gsim 6-7$ au), the near constancy of $T_J$ breaks down as the perturbations of Saturn and the other outer planets become more important, so we cannot consider the restricted three-body problem (Sun-Jupiter-object) any longer. Despite this shortcoming, we find that $T_J$ is still a useful tool to discriminate between different dynamical behaviors, so we will use it in the following for all the considered range of $q$.

\section{The numerical Integrations}

We carried out numerical integrations with the Bulirsch-Stoer algorithm of the Mercury package \citep{Cham99}. We considered the motion of massless bodies in a heliocentric frame under the gravitational influence of the Sun and seven planets (from Venus to Neptune), while the mass of Mercury was thrown into the Sun.  We used an accuracy parameter of $10^{-12}$.

\subsection{Short-term dynamical integrations}
 
In order to learn about the orbit evolution of the sample objects, we integrated the orbits of every Centaur with 50 clones backwards in time for $5 \times 10^4$ yr or until the tight bundle of the nominal orbit together with those of the 50 clones started to get loose and diverge, indicating that chaos made any prognosis of the orbit state further back in time uncertain. We will call this time $t_{div}$. All the Centaurs' orbits of our sample became chaotic well before reaching the integration time limit, so in general $t_{div} << 5 \times 10^4$ yr. The clones were generated by taking their orbital elements randomly within intervals $p \pm \sigma_p/2$, where $p$ is the nominal value of a given orbital element and $\sigma_p$ its uncertainty as provided by the JPL Solar System Dynamics Database. The typical uncertainties of the nominal values of the orbital elements are: semimajor axis between $10^{-6}$ and $10^{-3}$ au for orbits with quality codes between 0 and 5, eccentricities between $10^{-7}$ and $10^{-4}$, and for the angular parameters uncertainties between $10^{-5}$ and $10^{-2}$ degrees, all of these again for quality codes between 0 and 5.

\subsection{Long-term dynamical integrations}

When we extend the integrations to time scales $>> 10^4$ yr, we cannot expect to obtain accurate orbital determinations, so our results will have only a statistical validity and we need to follow the evolution with a high number of clones. We then generated for this study 299 clones for each observed Centaur, so we produced a sample of $68 \times 300 = 20,400$ inactive objects and $55 \times 300 = 16,500$ active objects. The clones were generated in the same manner as explained before for the short-term integrations.

The objects were integrated for 10 Myr into the future. If a test body reached a heliocentric distance of 1000 au the integration was terminated and the body was assumed to be ejected.

We assumed that a ``collision'' with the Sun occurred if the object reached a perihelion distance smaller than the Roche radius given approximately by

\begin{equation}\label{rtide}
r_{Roche} \simeq 2.5\left(\frac{\rho_{\odot}}{\rho_c}\right)^{1/3} R_{\odot} \simeq 0.012 \mbox{ au,}
\end{equation}
where $\rho_{\odot}$ and $\rho_c$ are the bulk densities of the Sun and the Centaur, and $R_{\odot}$ is the Sun's radius. We assumed for the Centaur a bulk density $\rho_c = \rho_{\odot}$.

\section{The results}

\subsection{The recent past of the Centaur population}

From the short-term numerical integrations we define the following parameter: $\nu = (q_{div} - q_o) / \Delta t$ (expressed in au yr$^{-1}$) where $\Delta t = t_o - t_{div}$ is the time elapsed between the present time $t_o$ and the time $t_{div}$ at which the particle starts to lose memory of its past evolution. $q_{div}$ and $q_o$ are the perihelion distances of the Centaurs at $t_{div}$ and $t_o$ respectively. $\nu$ gives us a measure of the rate at which $q$ changed in the recent past (some hundreds to thousands years). Positive $\nu$ values indicate that $q$ dropped from higher values, while negative $\nu$ values indicate a raise in $q$.

\begin{figure}[h]
\resizebox{10cm}{!}{\includegraphics{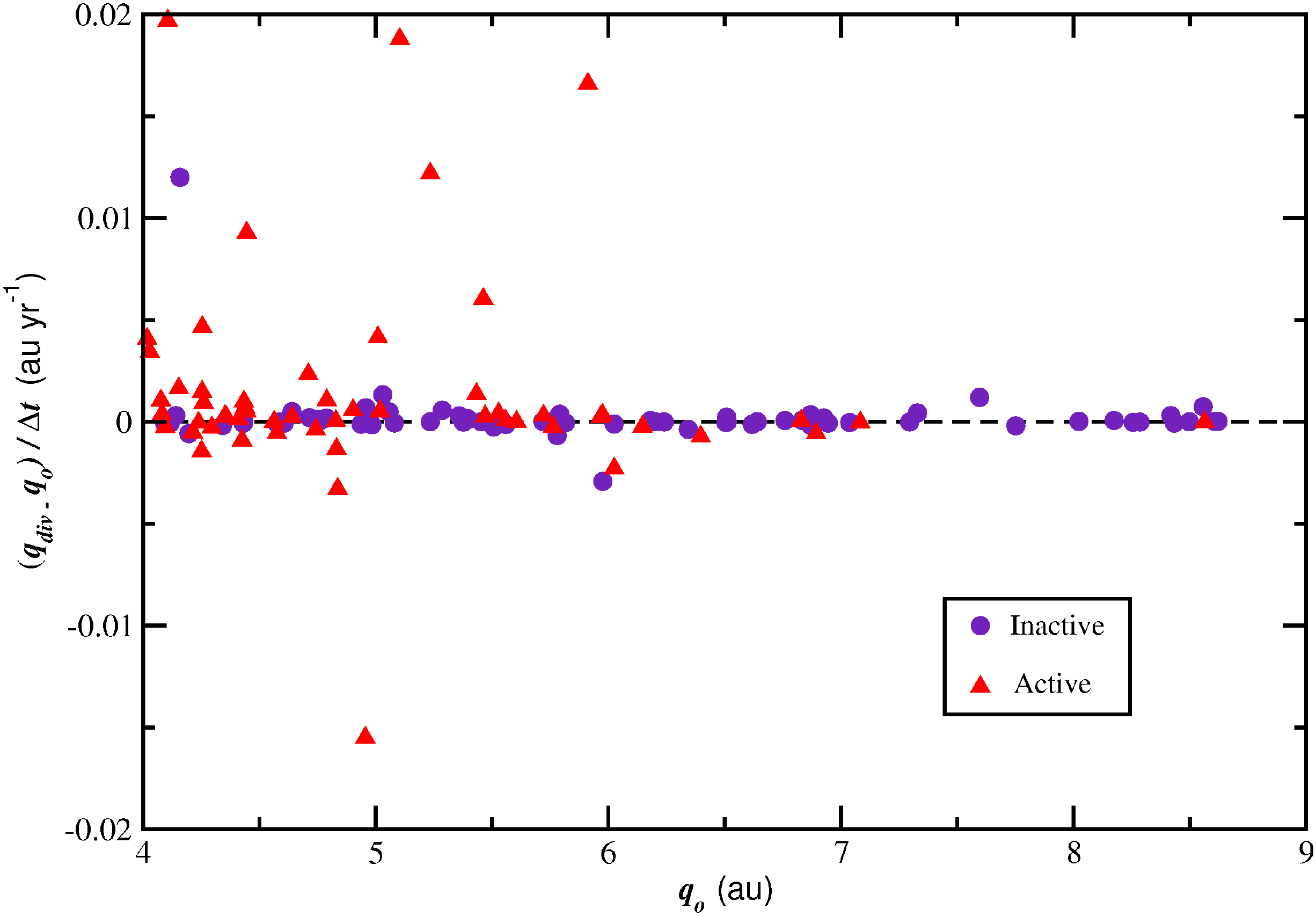}}
\caption{Rate of change of the perihelion distance in the recent past (time scales $\sim 10^2-10^3$ yr) as a function of the current perihelion distance for active and inactive Centaurs.}
\label{Centauros_Memoria}
\end{figure}

In Fig. \ref{Centauros_Memoria} we plot the rate of change $\nu$ as a function of the current perihelion distance of the Centaurs. We can see that the great majority of the inactive Centaurs stay very close to $\nu = 0$, indicating very little change in the recent past, while the spreading of values (mainly toward the positive side) is more evident for active Centaurs.

\subsection{Dynamical lifetimes of the Centaur populations and end states}

\begin{figure}[h]
\resizebox{12cm}{!}{\includegraphics{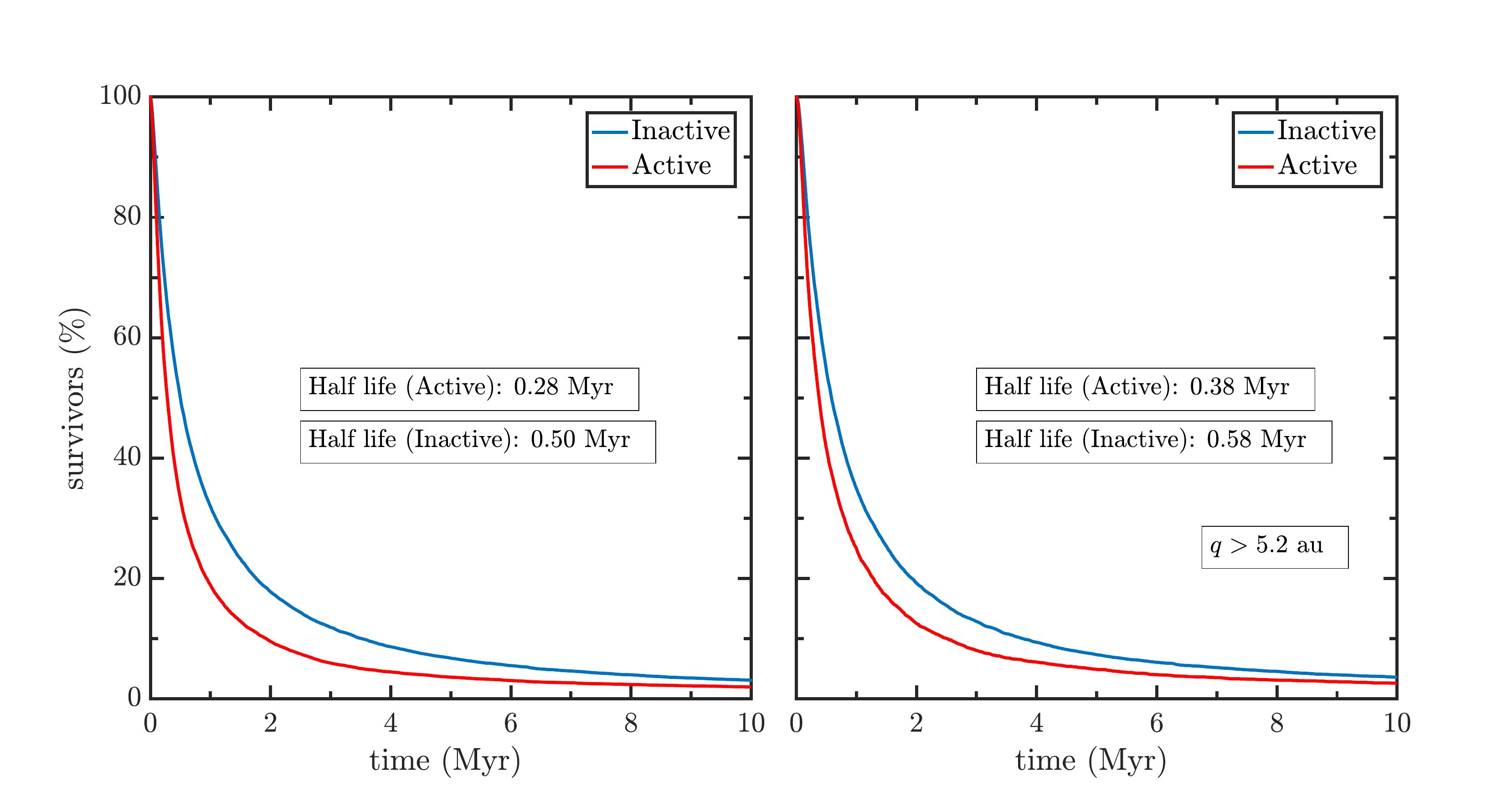}}
\caption{Percentage of survivors during the integration for 10 Myr for active and inactive Centaurs. We show results for the whole samples (left panel) and the samples restricted to bodies with $q > 5.2$ au (right panel).}
\label{Centaurs_life}
\end{figure}

For the long-term integrations, we analyze first the lifetimes of the two populations. We show in Fig. \ref{Centaurs_life} (left panel) the decay of the populations of active and inactive Centaurs as a function of time. The median dynamical lifetime or half-life (i.e. the time required for a sample to decay to a half) of inactive Centaurs is found to be nearly twice longer than that of active Centaurs. This can be explained because the latter have orbits on average closer to Jupiter and also have lower inclinations, which increases the probability of strong interactions with this planet.

To check whether the consideration of bodies in Jupiter-crossing orbits introduces a strong bias in the results, we also computed the half-lives of the more distant Centaurs with $q > 5.2$ au (Fig. \ref{Centaurs_life}, right panel). We can see that the half-life of inactive Centaurs is still about 50\% longer than that for active Centaurs.

A summary of the end states attained by the active and inactive Centaurs is shown in Table 1.

\begin{table}[h]
\centerline{Table 1: End states of our samples of Centaurs (\%)}
\begin{center}
\begin{tabular}{l l l} \hline
 End state & Inactive & Active\\ \hline
 Sungrazers & 3.03 & 0.93 \\
 Collision with Jupiter & 0.89 & 1.35 \\
 Collision with Saturn  & 0.14 & 0.16 \\
 Ejected & 92.83  & 95.57 \\
 Survivors at 10 Myr & 3.09 & 1.98 \\ \hline
\end{tabular}
\end{center}
\end{table}

\subsection{Sungrazers and collisions with the planets}

We show in Table 1 the percentage of test bodies that physically collide with Jupiter and Saturn. We find a somewhat greater percentage of Jupiter colliders among the active Centaurs because their inclinations are on average lower. Yet the percentage of Saturn colliders is similar for active and inactive Centaurs suggesting that Saturn has in general little influence on the dynamics of both populations. We did not find physical collisions with the terrestrial planets. To estimate the probability of a physical collision with the Earth, we computed close encounters with the Earth within one Hill radius and extrapolated the frequency of encounters down to what should be expected within the Earth's gravitational radius of collision, bearing in mind that the frequency of encounters is proportional to the square of the target radius. The gravitational radius of collision of a planet is given by

\begin{equation}\label{collis}
R_G = R\left(1 + \frac{v_{esc}^2}{u^2}\right)^{1/2}
\end{equation}
where $R$ is the planet's radius, $v_{esc}$ is the escape velocity of the planet and $u$ the encounter velocity of the body with the planet. In our case, the mean encounter velocity for the active Centaurs is found to be $<u> = 18.7$km/s, while we obtained $<u> = 27.1$ km/s for inactive Centaurs. These are rather high values ($> v_{esc} \simeq 11.2$ km/s for the Earth), so $R_G$ is only slightly greater than $R$.  

\begin{figure}[h]
\resizebox{14cm}{!}{\includegraphics{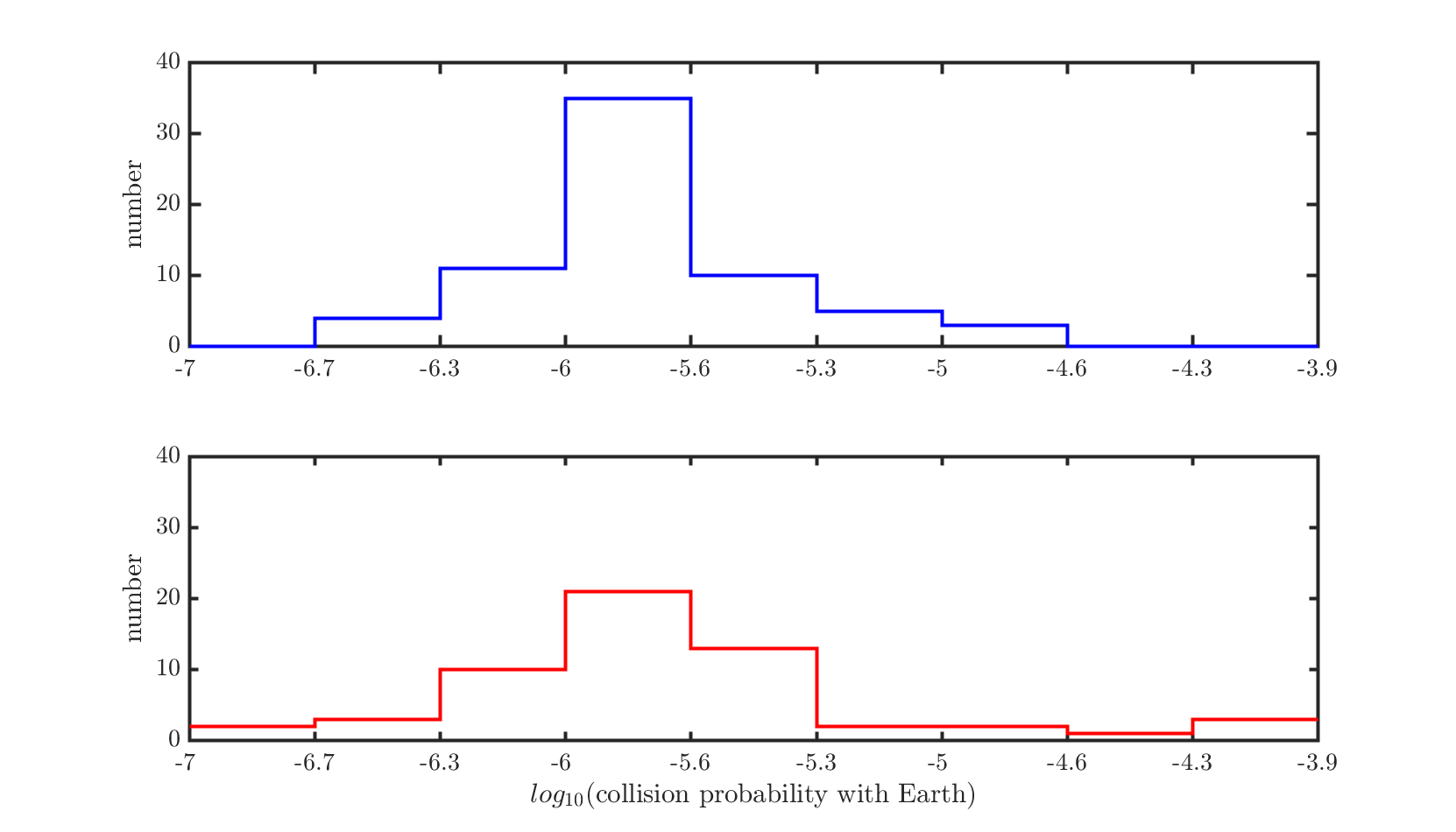}}
\caption{Histograms showing the distribution of probabilities that a given object will collide with the Earth during its evolution for both inactive (above) and active Centaurs (below).}
\label{collision_earth}
\end{figure}

We show in Fig. \ref{collision_earth} the distribution of probabilities of collision with the Earth for both, the active and inactive Centaurs. A typical value for both is around $2 \times 10^{-6}$, that can be interpreted as if we had $5 \times 10^5$ bodies moving in orbits similar to the considered Centaur, we should expect that one of them will collide with the Earth. Both inactive and active Centaurs are found to have similar chances to collide with the Earth though there are a few active Centaurs with greater probabilities of collision. We should bear in mind that these probabilities of collision do not take into account physical effects, for instance sublimation or thermal stresses, that might cause erosion, fragmentation and the possible disintegration of the body. The disintegration of a massive Centaur in the near-Earth region will produce a swarm of debris greatly increasing the probability of collision of some of them with the Earth \citep{Napi15}, though on the other hand preventing a rare but potentially deleterious megaimpact with global consequences for life and climate on Earth. 

Inactive Centaurs are more prone to become sungrazers than active ones. The reason may lie in their on average greater inclinations that favor the action of the Kozai mechanism, main responsible for bringing objects close to the Sun \citep{Bail92}.

\subsection{Evolution to Halley type and Jupiter family orbits}

\begin{figure}[h]
\begin{center}
\begin{tabular}{c}
\includegraphics[width=0.6\textwidth]{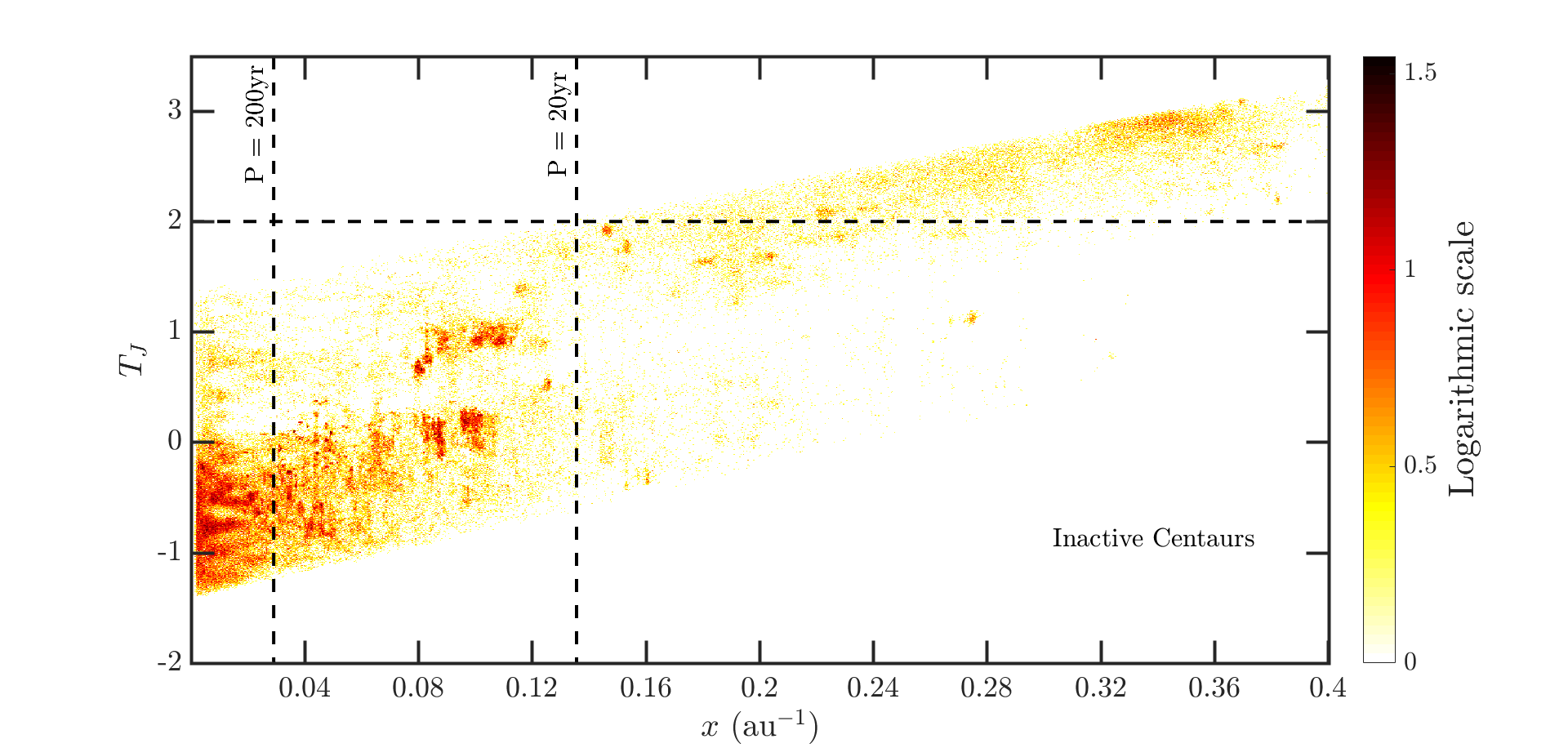}\\
\includegraphics[width=0.6\textwidth]{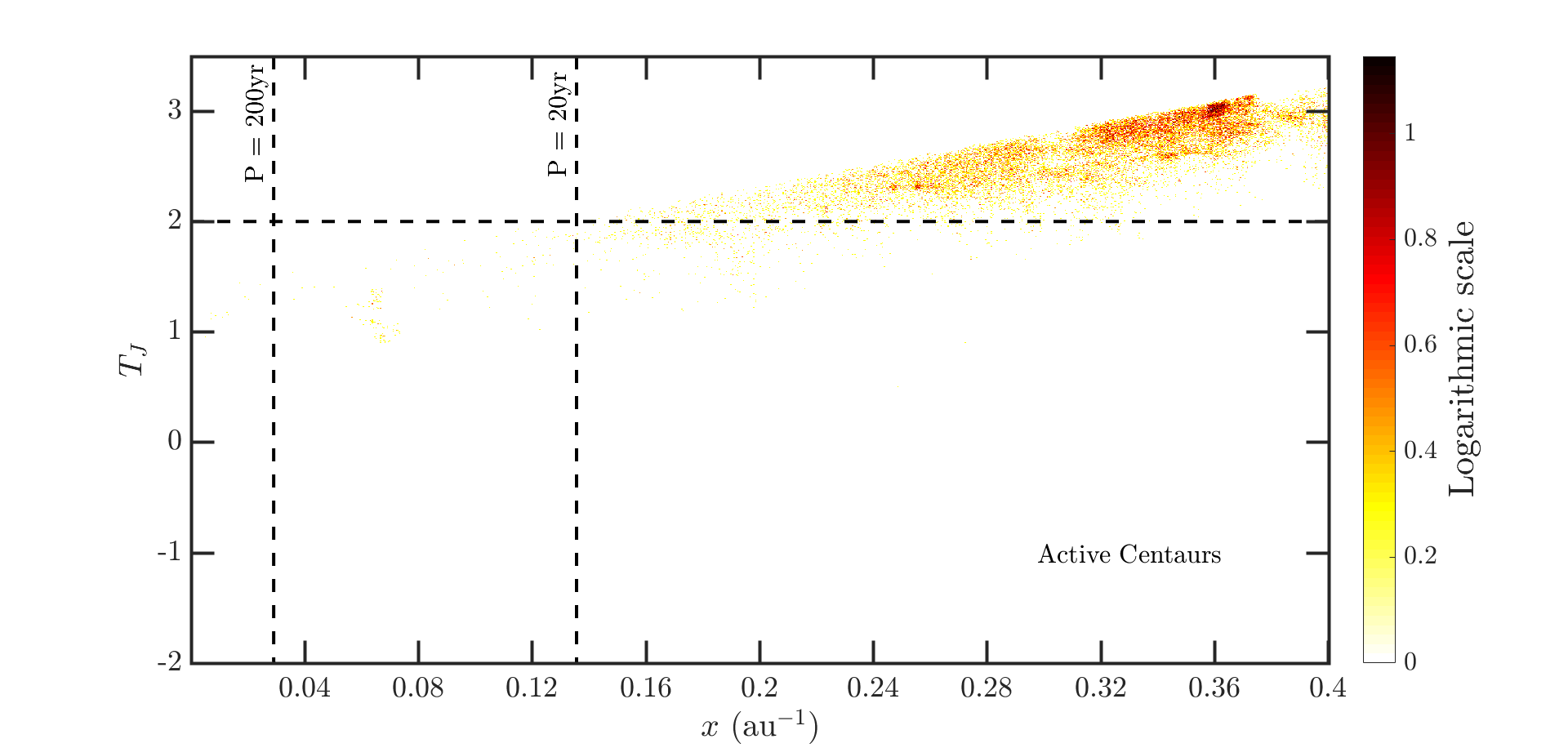}
\end{tabular}
\end{center}
\caption{Residence time in the plane Tisserand parameter versus energy of the computed inactive Centaurs that reach  $q<1.3$ au (upper panel). Idem for the active Centaurs (lower panel). The unit in the right-side bars is $10^3$ yr.}
\label{Centaurs_res_tvsoe}
\end{figure}

During the evolution, the perihelion distances of some Centaurs are scattered toward the inner planetary region where they become JFCs or HTCs. JFCs and some HTCs have very tightly bound orbital energies $x$. We note that the orbital energy $\propto 1/a$, so we adopt $x \equiv 1/a$ (expressed in au$^{-1}$) as a proxy of the energy. Therefore, the greater $x$, the smaller $a$ and the orbital period $P=x^{-3/2}$ yr. JFCs are conventionally defined as comets with orbital periods $P < 20$ yr (bound energies $x > 0.136$ au$^{-1}$), while HTCs extend conventionally to $P=200$ yr (bound energy $x = 0.0292$ au$^{-1}$). Comets with $P > 200$ yr are defined as of ``long-period''. We show in Fig. \ref{Centaurs_res_tvsoe} the residence times in the plane $(x,T_J)$ of the near-Earth objects ($q < 1.3$ au) that are produced in our samples of inactive and active Centaurs and their clones. While inactive Centaurs lead to Earth-crossers that spread in a wide range of energies and Tisserand parameters (Fig. \ref{Centaurs_res_tvsoe}, upper panel), active Centaurs lead to Earth-crossers heavily concentrated in the upper-right corner of Fig. \ref{Centaurs_res_tvsoe} (lower panel) within the ranges $0.136 < x < 0.4$ au$^{-1}$ (orbital periods $4 \lsim P < 20$ yr) and $2 < T_J < 3$.

\begin{figure}[h]
\resizebox{12cm}{!}{\includegraphics{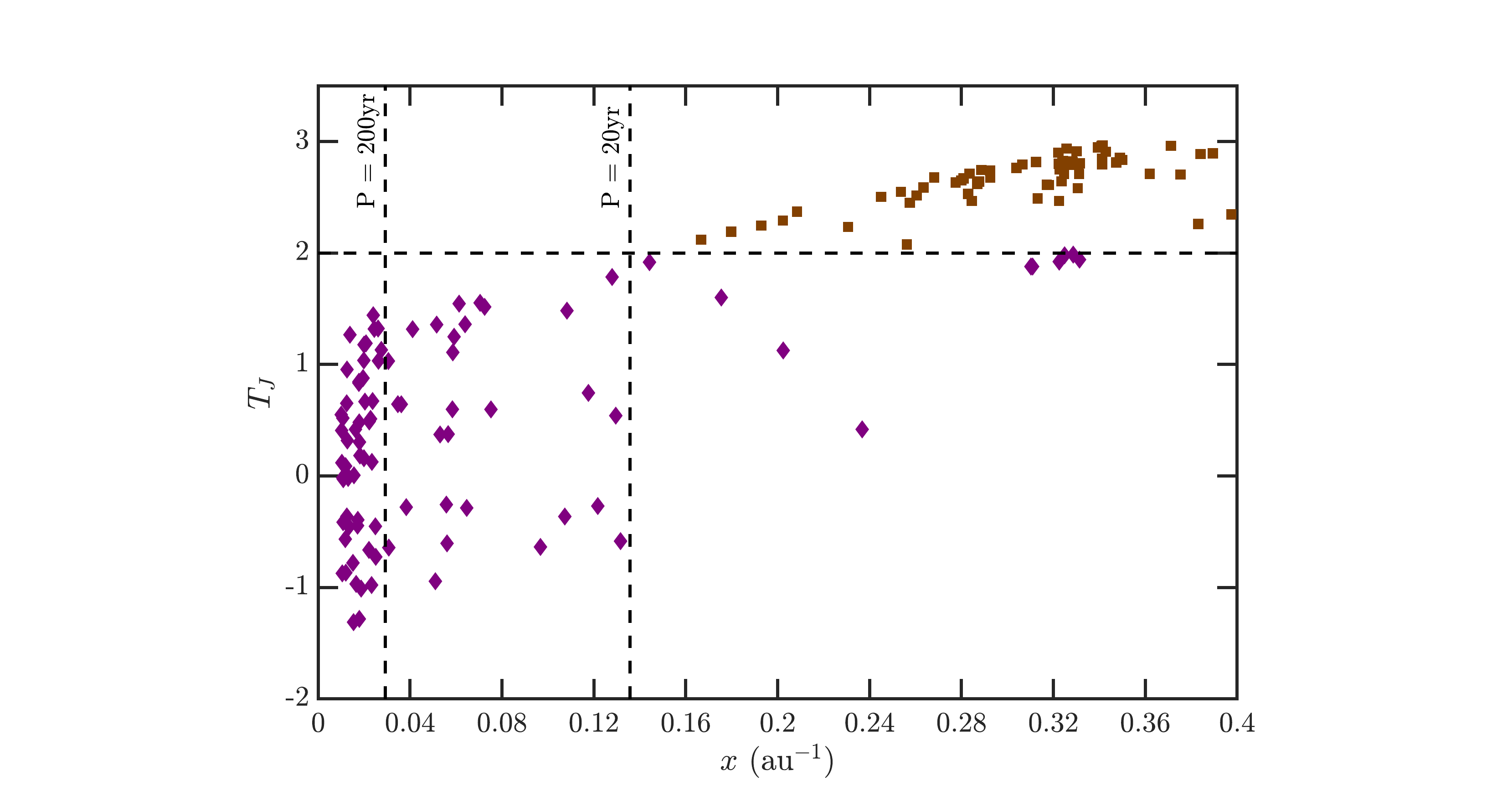}}
\caption{Observed comets with $q<1.3$ au and orbital periods $P < 10^3$ yr plotted in the plane Tisserand parameter versus energy.}
\label{Centaurs_obs}
\end{figure}

In Fig.\ref{Centaurs_obs} we show a plot of the energy versus the Tisserand parameter for the observed comets with $q<1.3$ au and orbital periods $P < 10^3$ yr ($x > 10^{-2}$ au$^{-1}$), namely a sample comprising JFCs, HTCs and dynamically old LPCs. Note that JFCs have values $T_J > 2$ while HTCs and old LPCs have values $T_J  < 2$. By comparing the observed samples with the computed ones, we find a clear mismatch between the observed samples and the computed samples taken separately, namely inactive Centaurs produce essentially HTCs and old LPCs, but very few JFCs. On the other hand, active Centaurs produce essentially JFCs but very few HTCs. If we combined the computed JFCs and HTCs produced by both inactive and active Centaurs, we then get a good match with the observations.

\begin{figure}[h]
\begin{center}
\begin{tabular}{c}
\includegraphics[width=0.6\textwidth]{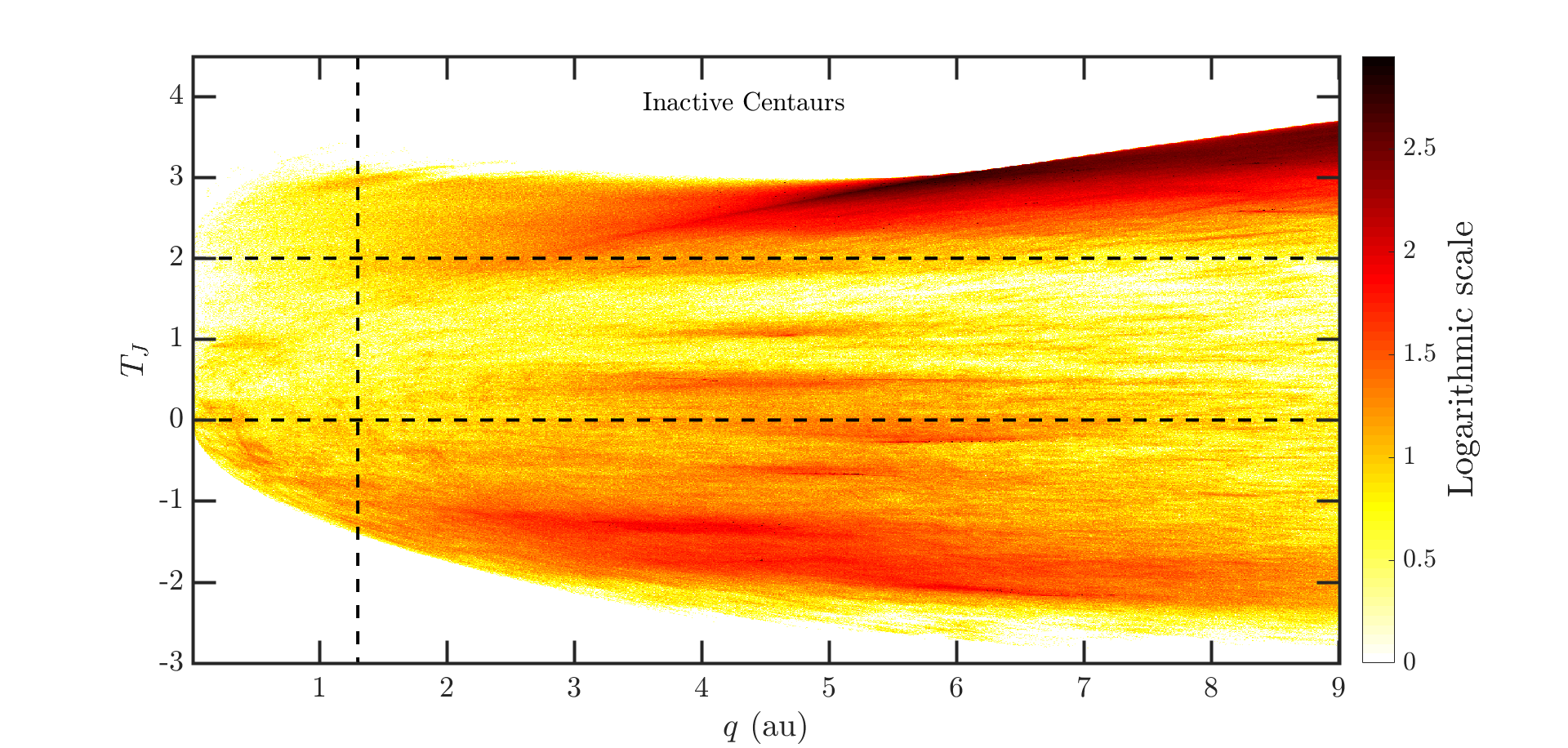}\\
\includegraphics[width=0.6\textwidth]{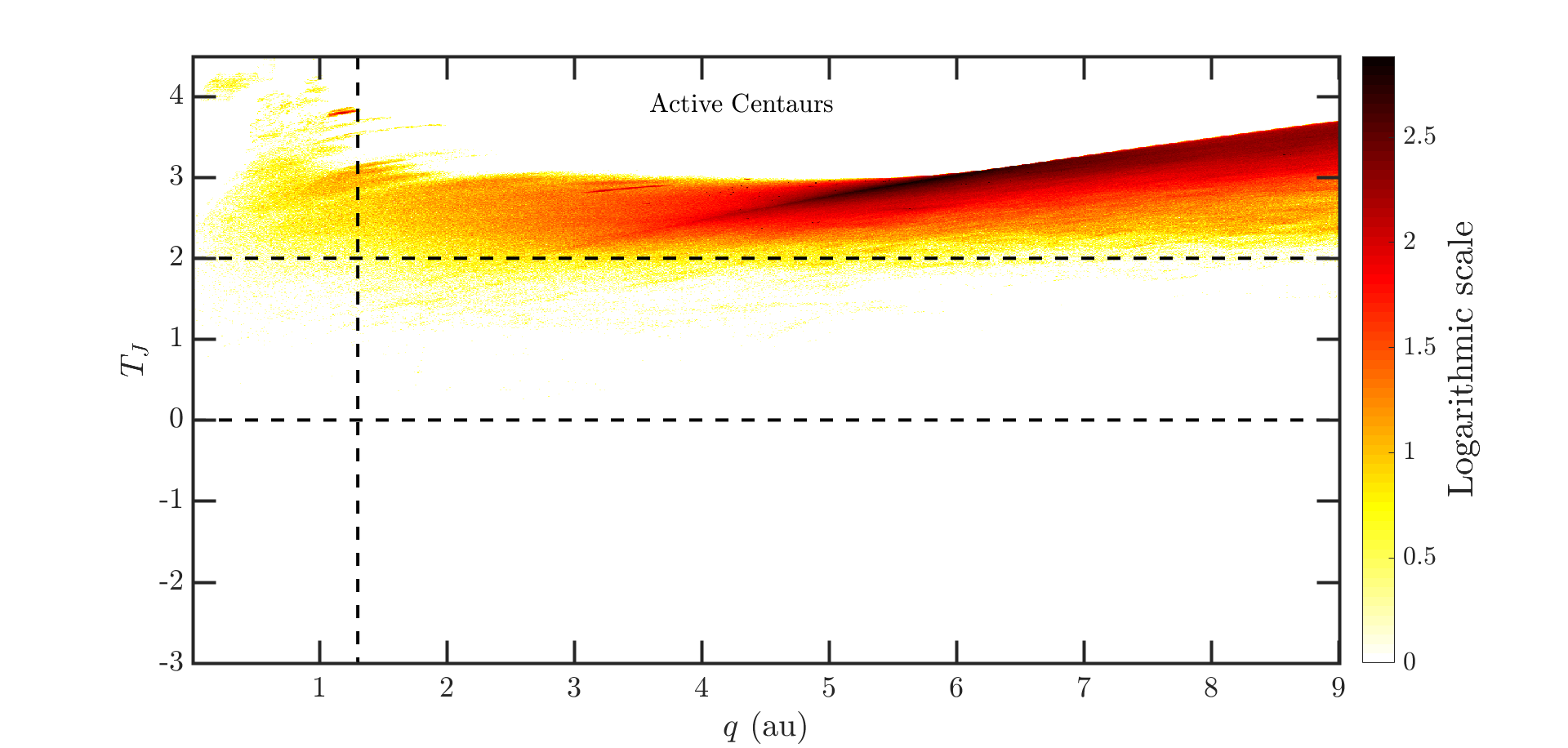}
\end{tabular}
\end{center}
\caption{Residence time in the plane perihelion distance versus Tisserand parameter of the computed inactive (top) and active (bottom) Centaurs. We see that the time spent with $q < 1.3$ au represents only a very small fraction of the total residence time in the region of interest for both active and inactive Centaurs. The unit in the right-side bars is $10^3$ yr.}
\label{Centaurs_res_tvsq}
\end{figure}

In Fig. \ref{Centaurs_res_tvsq} we show the residence times in the plane $(q,T_J)$ for both populations during the integration. The dynamical evolution proceeds from right to left keeping $T_J$ nearly constant, even for those Centaurs starting with $q \gsim 6$ au, which supports the pertinence of using $T_J$ as a quasi-invariant of the evolution even in cases where the Centaurs do not come very close to Jupiter. Some Centaurs with initial $q \gsim 5.5$ au have $T_J$ above three, but the Tisserand parameter decreases to values around three or somewhat below as soon as the objects approach or cross Jupiter's orbit.

As expected, active Centaurs evolve toward the Sun confined within a strip in Tisserand parameters $2 \lsim T_J \lsim 3$ (except for a few bodies perturbed by the terrestrial planets into orbits of small semimajor axes whose $T_J$ values raise above three). On the other hand, inactive Centaurs evolve within a much broader strip covering the range $-2.5 \lsim T_J \lsim 3$.

\begin{figure}[h]
\resizebox{12cm}{!}{\includegraphics{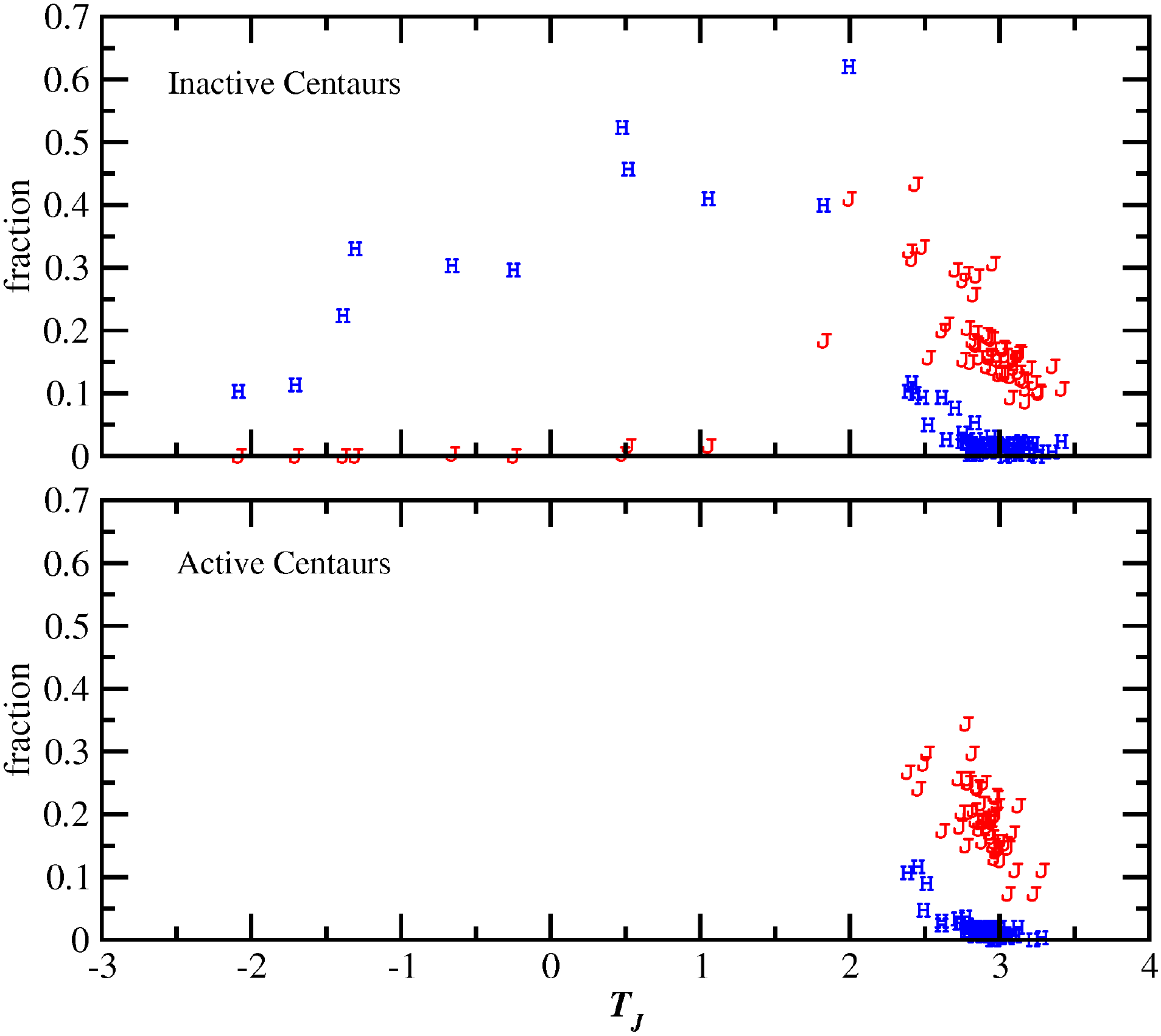}}
\caption{Fraction of clones of our samples of active and inactive Centaurs that reach HTC (H) and JFC orbits (J) with $q < 1.3$ au as a function of the initial Tisserand parameter of the Centaurs.}
\label{tj_htc_jfc}
\end{figure}

Figure \ref{tj_htc_jfc} plots the likelihood of a Centaur to become an HTC or JFC in near-Earth orbit ($q < 1.3$ au). This has been computed as the number of clones of a given Centaur that reach a near-Earth orbit, either as a HTC or as a JFC, divided by the total number of clones of the object (= 300). The figure shows that HTC-prone objects have initial Tisserand parameters with respect to Jupiter $T_J \lsim 2.4$ and are all inactive Centaurs, while JFC-prone objects have $T_J \gsim 2.4$ and are all the active Centaurs plus a fraction of the inactive Centaurs. Near-Earth HTCs are found to have a median dynamical lifetime of $1.75 \times 10^4$ yr, while the corresponding value for JFCs is $5 \times 10^3$ yr, in good agreement with some previous estimates \citep{Disi09}.

\subsection{Resonant states during the evolution}

\begin{figure}[h]
\resizebox{12cm}{!}{\includegraphics{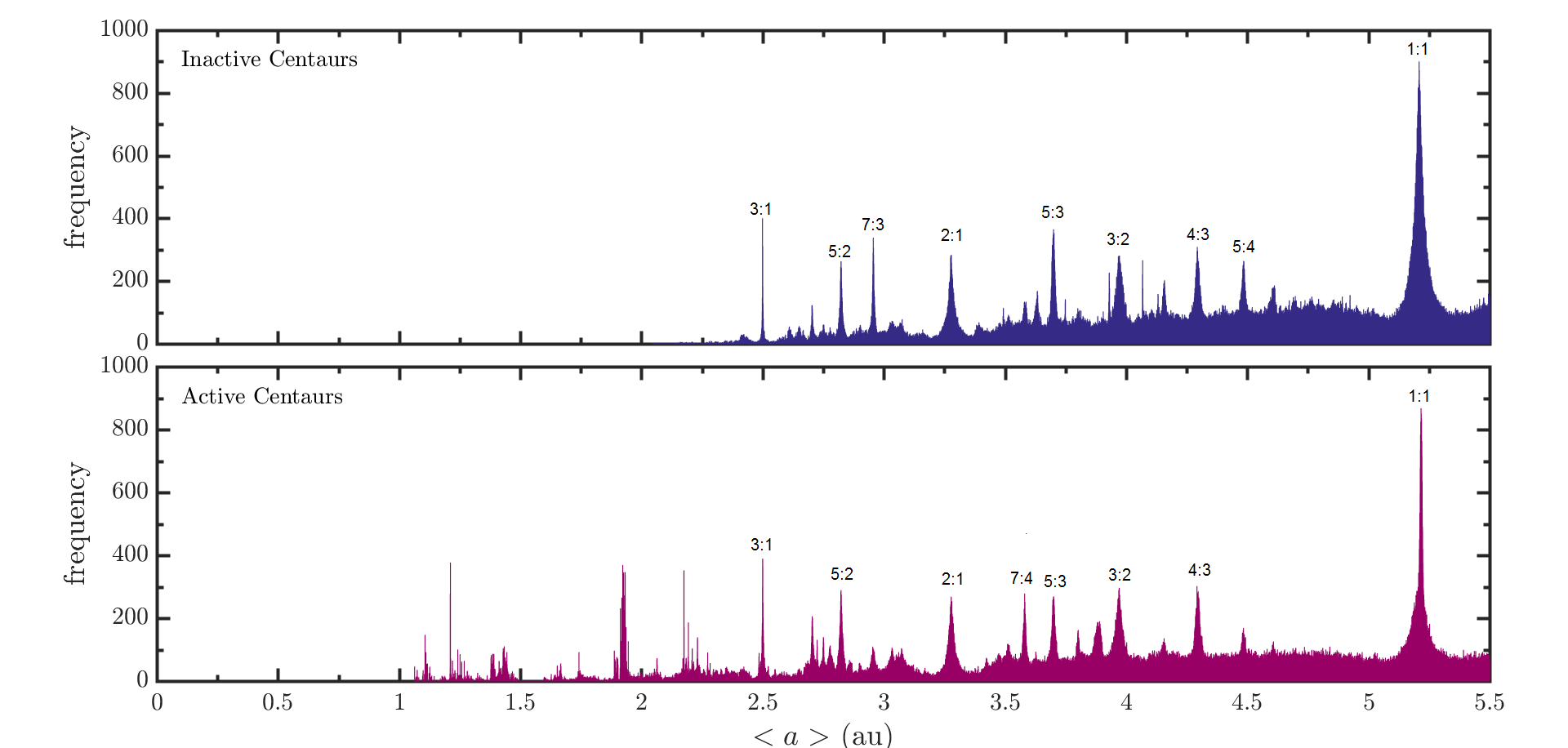}}
\caption{Histogram of the semimajor axis averaged over $10^4$ years for both inactive and active Centaurs that reach semimajor axes $a < 5.5$ au. Mean motion resonances with Jupiter are labeled.}
\label{Centaurs_resonances}
\end{figure}

Another interesting aspect of the dynamical evolution is resonance capture. Averaged semimajor axes are good indicators of resonant orbital states. In order to study the relevance of the MMRs we have to eliminate short-period oscillations in the orbital elements, so we average the orbital elements in time intervals of $10^4$ yr. In Fig.\ref{Centaurs_resonances} we have a histogram of $<a>$ for each population.

Some of the resonances that can be easily identified are the Hildas (3:2) at $a = 3.98$ au, the Thules (4:3) at $a = 4.29$ au, the 2:1 at $a=3.28$ au, the 3:1 at $a=2.50$ au and the co-orbital bodies of Jupiter (1:1) at $a = 5.2$ au that include Trojans. This suggests a possible mechanism for re-populating these resonances with Centaurs. However, we stress the fact that the orbits of these new members are not stable.

To have an idea of the stability of the orbits we show in Table 2 some statistics regarding the residence times $t_r$ of test bodies in some of the principal mean motion resonances. We also show the total number of inactive Centaurs and their clones (top) and active Centaurs and their clones (bottom) that reach each one of the resonances for time spans $>20$ kyr. For instance, the probability that any of the inactive Centaurs (with their clones = 20400) will fall in a co-orbital state for time spans $> 20$ kyr is 372/20400 $\simeq$ 0.018, while for active Centaurs (with their clones = 16500) is 271/16500 $\simeq$ 0.016. We stress that these are average probabilities for the ensembles of active and inactive Centaurs (and their clones). We note that a large fraction of the test bodies are ejected before reaching the co-orbital state. Trapping as a Jupiter's co-orbital body is the most common state that does not necessarily imply a Trojan state, it can also involve horseshoe trajectories and quasi-satellites of Jupiter.

\begin{table}[h]
  \centerline{Table 2: Distribution of times spent in different resonances}

  \centerline{for time spans $t_r > 20$ kyr}

  \centerline{}

\begin{center}
  \begin{tabular}{l r r r r r} \hline
    & & Inactive Centaurs & & \\ \hline
    Family & No. Clones & 20-30 kyr & 30-100 kyr & $> 100$ kyr \\
    & & (\%) & (\%) & (\%) \\ \hline
    Co-orbital (1:1) & 372 & 50.48 & 42.31 & 7.21 \\
    Thule (4:3) & 240 & 37.99 & 48.53 & 13.48 \\
    Hilda (3:2) & 181 & 50.74 & 34.48 & 14.78 \\
    (2:1)  & 74  & 37.23  & 47.87 & 14.89 \\
    (3:1)  & 6  & 40.91  & 45.45 & 13.64 \\ \hline
     & & Active Centaurs & & \\ \hline
    Family & No. Clones & 20-30 kyr & 30-100 kyr & $> 100$ kyr \\
    & & (\%) & (\%) & (\%) \\ \hline
    Co-orbital (1:1) & 271 & 51.37 & 44.20 & 4.44 \\
    Thule (4:3) & 259 & 43.54 & 45.63 & 10.83 \\
    Hilda (3:2) & 164 & 46.59 & 38.07 & 15.34 \\
    (2:1)  & 90  & 41.90  & 48.10  & 10.00 \\
    (3:1)  & 15  & 37.21  & 41.86 & 20.93 \\ \hline
\end{tabular}
\end{center}
\end{table}

By comparing the frequency of $<a>$ in the spikes with that in the background (the bottom colored region of Fig. \ref{Centaurs_resonances}), we find that active and inactive Centaurs reaching semimajor axes $a < 5.5$ au spend about 25\% of their evolution time inside interior two-body MMRs with Jupiter and this percentage holds for both populations. We show in Table 3 a summary of the contribution of the most important resonances.

\begin{table}[h]

  \centerline{Table 3: Percentage of the evolution time in some resonant states}

  \centerline{for Centaurs (and clones) that reach $a < 5.5$ au}
\begin{center}
  \begin{tabular}{l r r} \hline
 Resonant state & Inactive (\%) & Active (\%) \\ \hline
 Co-orbital (1:1) & 12.25 & 8.17 \\
 Thule (4:3) & 1.81 & 2.15 \\
 Hilda (3:2) & 2.27 & 2.79 \\ \hline
\end{tabular}
\end{center}
\end{table}

\begin{figure}[h]
\resizebox{12cm}{!}{\includegraphics{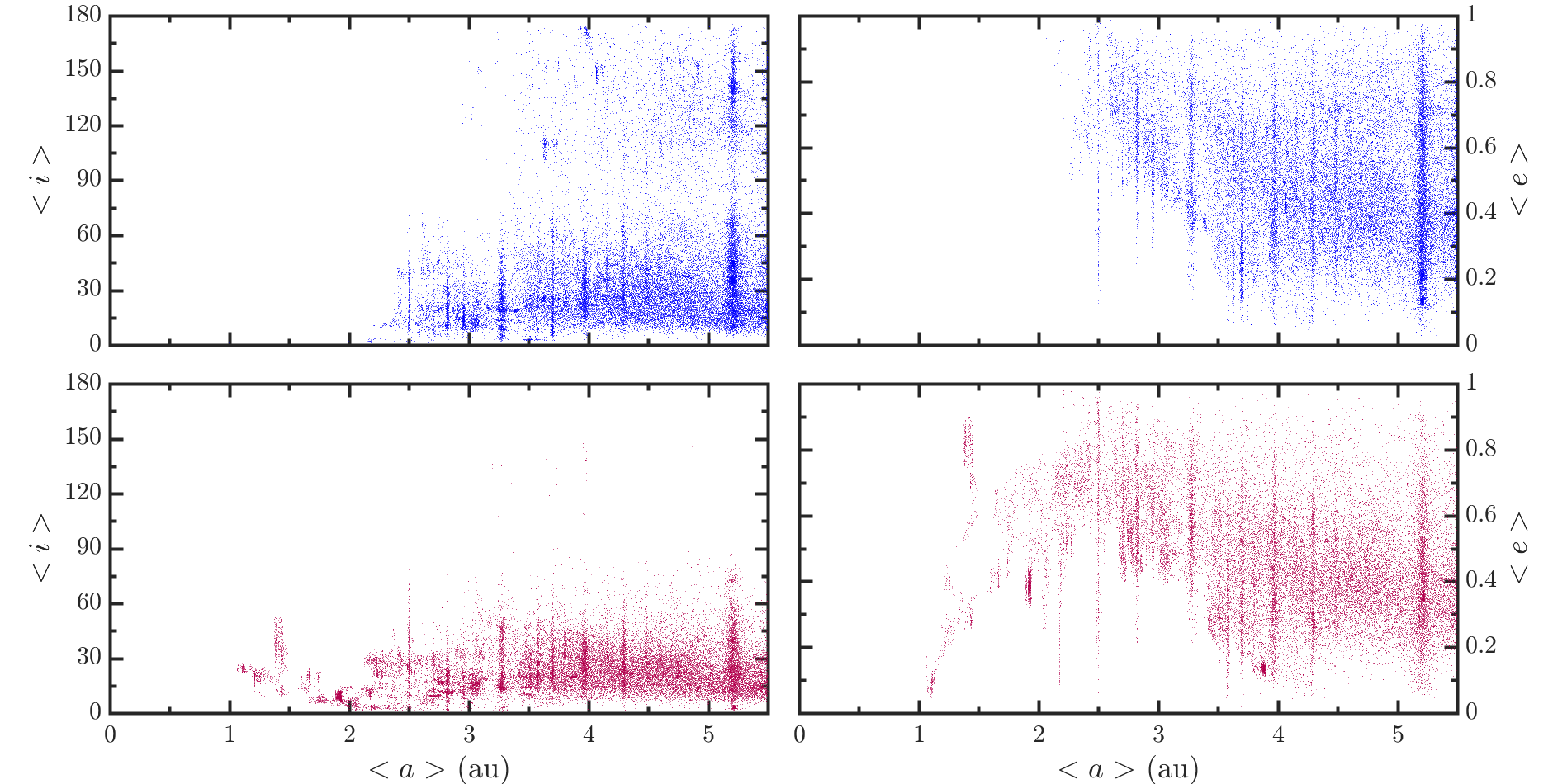}}
\caption{Evolution of the inactive Centaurs (upper panels), and the active Centaurs (lower panels) in the planes: mean semimajor axis ($<a>$) versus mean inclination ($<i>$) and mean semimajor axis ($<a>$) versus mean eccentricity ($<e>$).}
\label{Centaurs_resonance_car}
\end{figure}

In Fig.\ref{Centaurs_resonance_car} we plot the evolution of the computed Centaurs in the planes of the mean orbital elements $<a>$ versus $<i>$ and $<a>$ versus $<e>$ by taking snapshots every $10^3$ yr. Although we have resonances from highly eccentric to almost circular orbits, active Centaurs that move only on prograde orbits fall entirely in prograde resonances, while inactive Centaurs, that contain a fraction of retrograde orbits, fall in both prograde and retrograde resonances. In this regard we highlight the capture of quite a few retrograde Centaurs in retrograde co-orbital states with Jupiter (see upper-left panel of Fig. \ref{Centaurs_resonance_car}). In the last years retrograde resonances have been reported in some works and also motivated theoretical and numerical studies \citep{Namo15}, and it is also interesting to mention the discovery of object 2015 BZ$_{509}$, the first retrograde co-orbital asteroid of Jupiter \citep{Wieg17}.

Resonant objects undergo dynamical evolutions that force low eccentricity orbits, in particular the 3:1 MMR with Jupiter (Fig. \ref{Centaurs_resonance_car}, right panels). The consequence is that these bodies get decoupled from Jupiter during these resonant states. In fact, some of the objects captured in 3:1 MMR underwent large variations in eccentricity (from nearly zero to nearly one) and some of them end up as sungrazers. Thus, during the journey to the region of the terrestrial planets, Centaurs can become temporally interlopers in the main asteroid belt, in resonant as well as non-resonant states, and with moderate eccentricities ($e \sim 0.2-0.3$). Another interesting result is that the inactive population remains with $a>2$ au while the active one can reach $a<2$ au due to close encounters with the Earth and Venus. Of course, the latter may have a purely academic interest since finite physical lifetimes may greatly limit the possibility to attain such tightly bound orbits.

\begin{figure}[h]
\resizebox{12cm}{!}{\includegraphics{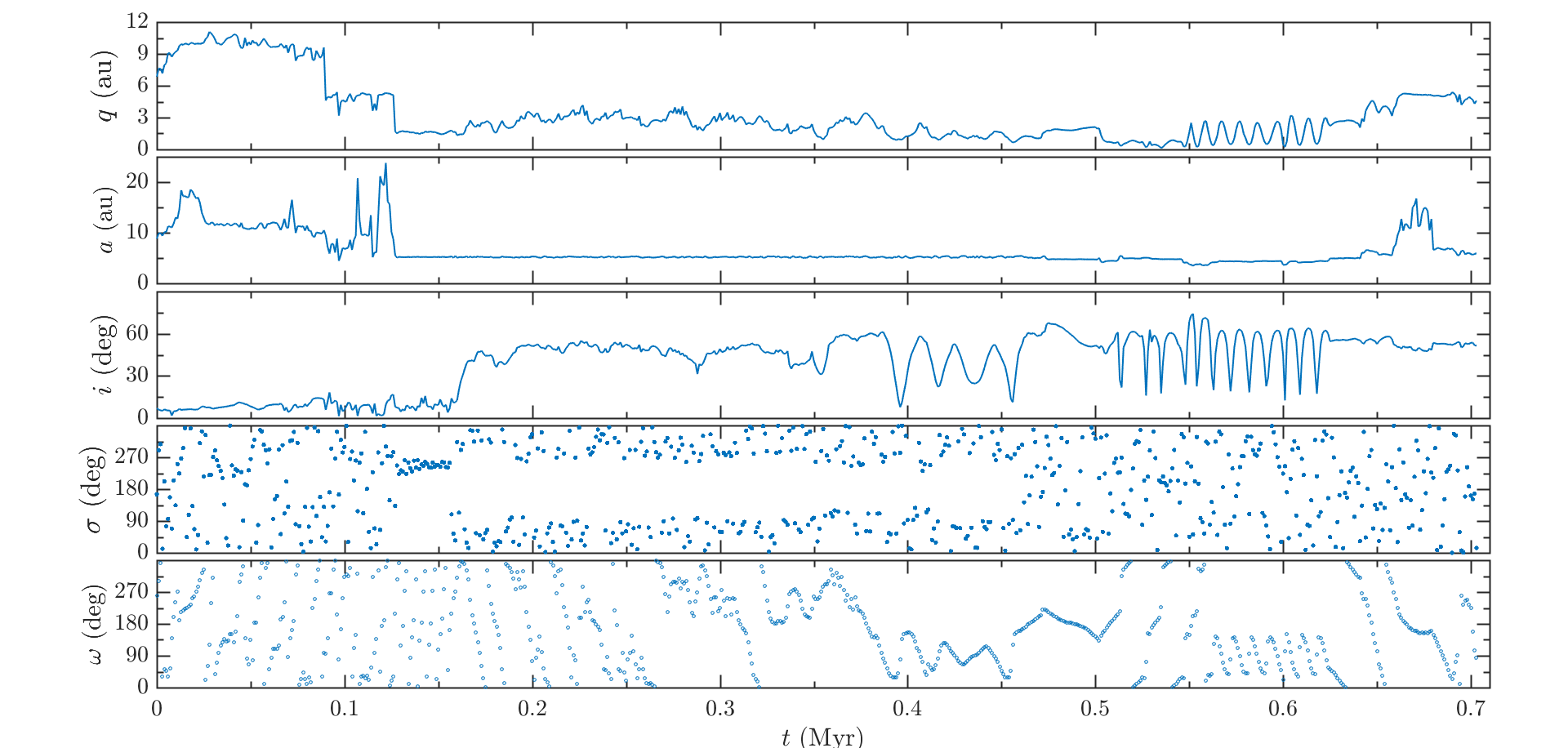}}
\caption{Evolution of one of the clones of the inactive Centaur 2013 VE2 as illustrated by the orbital parameters: perihelion distance ($q$), semimajor axis ($a$), inclination ($i$), critical angle ($\sigma$), and argument of perihelion ($\omega$).}
\label{Centaurs_resonance_2013VE2}
\end{figure}

Figure \ref{Centaurs_resonance_2013VE2} shows a clone of the inactive Centaur 2013 VE2 entering in the 1:1 resonant motion with Jupiter at $t=1.25 \times 10^5$ yr. The orbital evolution is depicted through the following parameters: $q$, $a$, $e$, the critical angle $\sigma$ and the argument of perihelion $\omega$. For a $k$-order resonance $|j+k|:|j|$, where $j$ and $k$ are integers, the critical angle is defined as

\begin{equation}\label{reson}
\sigma = (j+k)\lambda_p - j\lambda - k\varpi
\end{equation}
where $\lambda_p$, $\lambda$ are the mean longitudes of the planet and the object respectively and $\varpi$ is the longitude of perihelion. The angle $\sigma$ librates or has a slow time evolution in the resonance, while it circulates out of the resonance (e.g. Gallardo 2006). For a co-orbital state we have $j=-1$, $k=0$ so eq.(\ref{reson}) becomes $\sigma = \lambda - \lambda_p$. 

While in the 1:1 MMR, the clone of 2013 VE2 librates around the Lagrangean point L5 which, for the high eccentricity of $e \simeq 0.7$ and inclination $i \simeq 10^{\circ}$ attained by the test body at that moment, shifts to $\sigma \simeq 250^{\circ}$ \citep{Gall06}. Then, at $t=1.55 \times 10^5$ yr it switches to a quasi-satellite state and remains there up to $t=4.6 \times 10^5$ yr. In this state the critical angle librates around $\sigma = 0$. After leaving the co-orbital state with Jupiter, the object is captured by the Kozai mechanism during almost $10^5$ yr causing large coupled oscillations in $q$ and $i$. The object is ejected from the Solar System at $7.1 \times 10^5$ yr.

\section{Discussion}

Why do some Centaurs show activity while others not?, does it respond to chemical and/or physical differences in their constitution, perhaps related to different source regions?, or otherwise, it is due to different evolution histories? In this regard, the loss of ultrared matter is thought to be associated to a change in the color and albedo of the object, going from a bright red surface to a neutral ('gray') of lower albedo once the ultrared matter is lost. This loss is observed to occur at $\sim 10$ au, and this is more or less the perihelion distance at which activity starts to be observed which would suggest that the loss of the ultrared matter and the onset of activity are related \citep{Jewi15}. But this is part of the story since neutral-color (gray) Centaurs can be either active or inactive. Therefore, the loss of the ultrared matter does not warrant activity. We have to look into the properties of the subsurface material to find an explanation for the onset of activity.

The high inclinations of some inactive Centaurs might suggest a different source region perhaps in the Oort cloud \citep{Emel05,Bras12,Dela14}, but it is also possible that they come from the trans-neptunian belt, as the active Centaurs, with the difference that they are scattered to distances $\gsim 10^4$ au where galactic tides can drive their perihelia to the planetary region and force high inclinations \citep{Levi06}.

The HTC-prone objects are found only among inactive Centaurs. The active Centaurs (``comets'') tend to evolve to JFCs, though there is a small fraction of them that also spend some time as HTCs. Actually, we need both populations of active and inactive Centaurs to reproduce the near-Earth populations of JFCs and HTCs. Are therefore the inactive Centaurs the progenitors of most HTCs? If this is indeed the case, then the lack of activity observed at present may be due to the loss of surface/subsurface volatiles and/or the completion of the amorphous-crystalline phase transition in the subsurface layers after a long residence time in the region where they were discovered. We have to bear in mind that the dynamical time scale of inactive Centaurs is appreciably longer than that for active Centaurs. We should expect that inactive Centaurs will reactivate as soon as they decrease their perihelion distances to the point that the more intense Sun's radiation starts to sublimate volatiles underneath the dusty surface or the amorphous-crystalline phase transition of water ice proceeds at ever deeper layers. This may explain why inactive Centaurs lead to active HTCs.

\begin{figure}[h]
\resizebox{12cm}{!}{\includegraphics{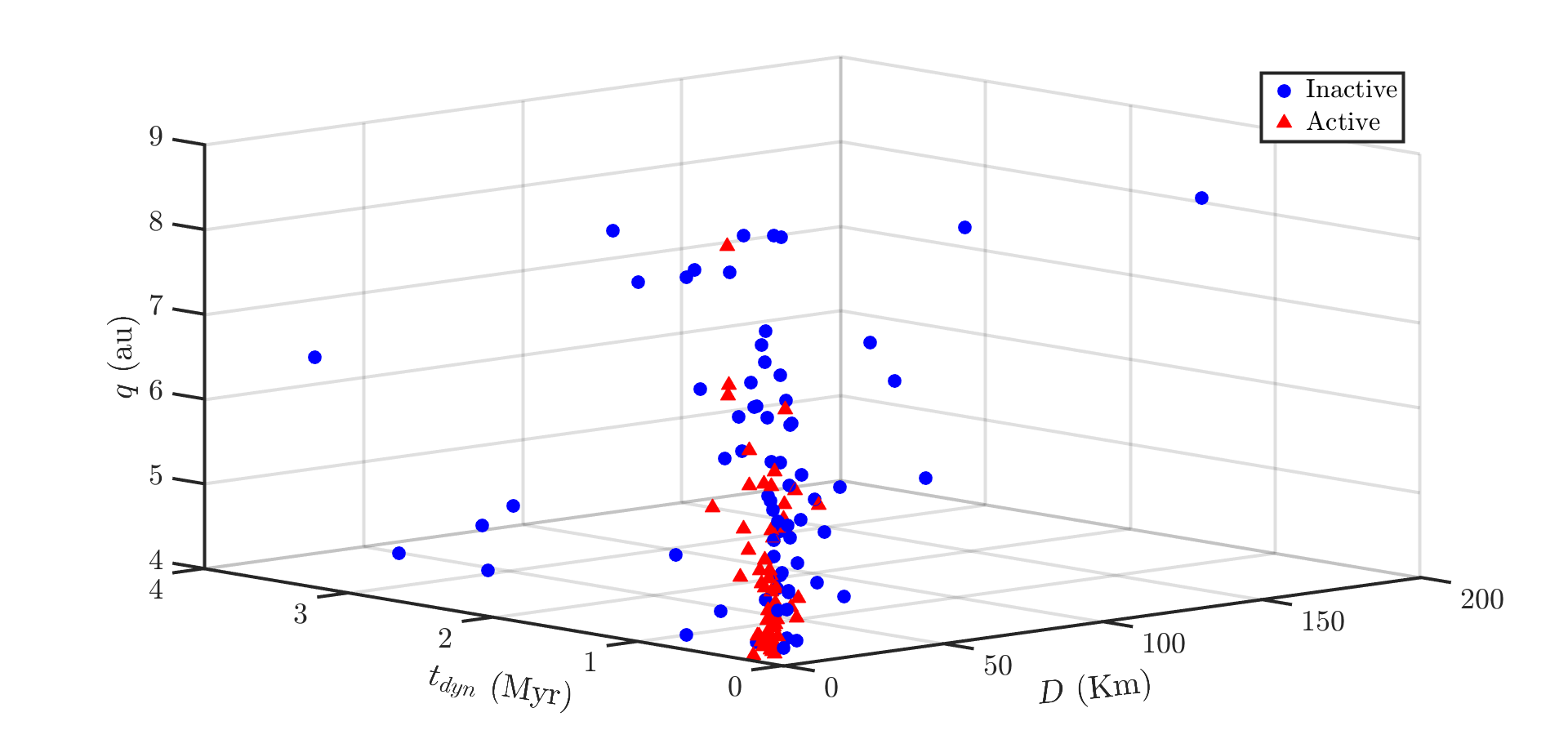}}
\caption{Distribution of active and inactive Centaurs in the parametric space Diameter - Dynamical lifetime - initial perihelion distance.}
\label{D_tdyn_q}
\end{figure}

We argue that most active and inactive Centaurs come from the same source region (the trans-neptunian region) but that they have experienced different dynamical evolutions. A larger permanence in the Jovian region may favor the deactivation of the object through the exhaustion of exposed highly volatile material (e.g. CO, CO$_2$), the completion of the amorphous-crystalline phase transition of water ice in subsurface layers, and/or the buildup of an insulating dust mantle. In this regard size may also play a role as larger bodies have more capacity to retain dust grains on the surface. Furthermore, the perihelion distance should also play a role as a closer approach to the Sun may trigger renewed activity \citep{Jewi09}.

We show in Fig. \ref{D_tdyn_q} the space distribution of active and inactive Centaurs in the parametric space: Object's diameter ($D$) - Dynamical lifetime ($t_{dyn}$) - initial perihelion distance ($q$). The diameter $D$ was derived from the absolute magnitude $H$ of inactive and active Centaurs taken from the JPL Solar System Dynamics and the MPC photometric databases. In the case of active Centaurs or comets we chose the fainter estimated magnitudes of each object that presumably approached the magnitude of the bare nucleus. To convert from $H$ to $D$ we used the expression \citep{Harr15}

\begin{equation}\label{d_vs_h}
  D = \frac{1329 \mbox{ km}}{\sqrt{p_v}} 10^{-H/5}
\end{equation}  
where $p_v$ is the (visual) geometric albedo. We adopted for inactive Centaurs an average $p_v = 0.08$ (cf. Bauer et al. 2013) (except for a few objects for which the albedo was known), and for the active Centaurs $p_v = 0.04$ corresponding to the value usually adopted for comets \citep{Snod11}. As regards the dynamical lifetime $t_{dyn}$ of a given Centaur, it was obtained as the average of the dynamical lifetimes of the object and its 299 clones.

We can see in the space $(D,t_{dyn},q)$ a clear segregation in the distribution of active and inactive Centaurs, such that the active ones tend to occupy the corner of smaller $D$, shorter $t_{dyn}$ and smaller $q$, strongly suggesting that activity is displayed when the body is dynamically young in the zone around Jupiter and is rather small (less than a few km).

\begin{figure}[h]
\resizebox{9cm}{!}{\includegraphics{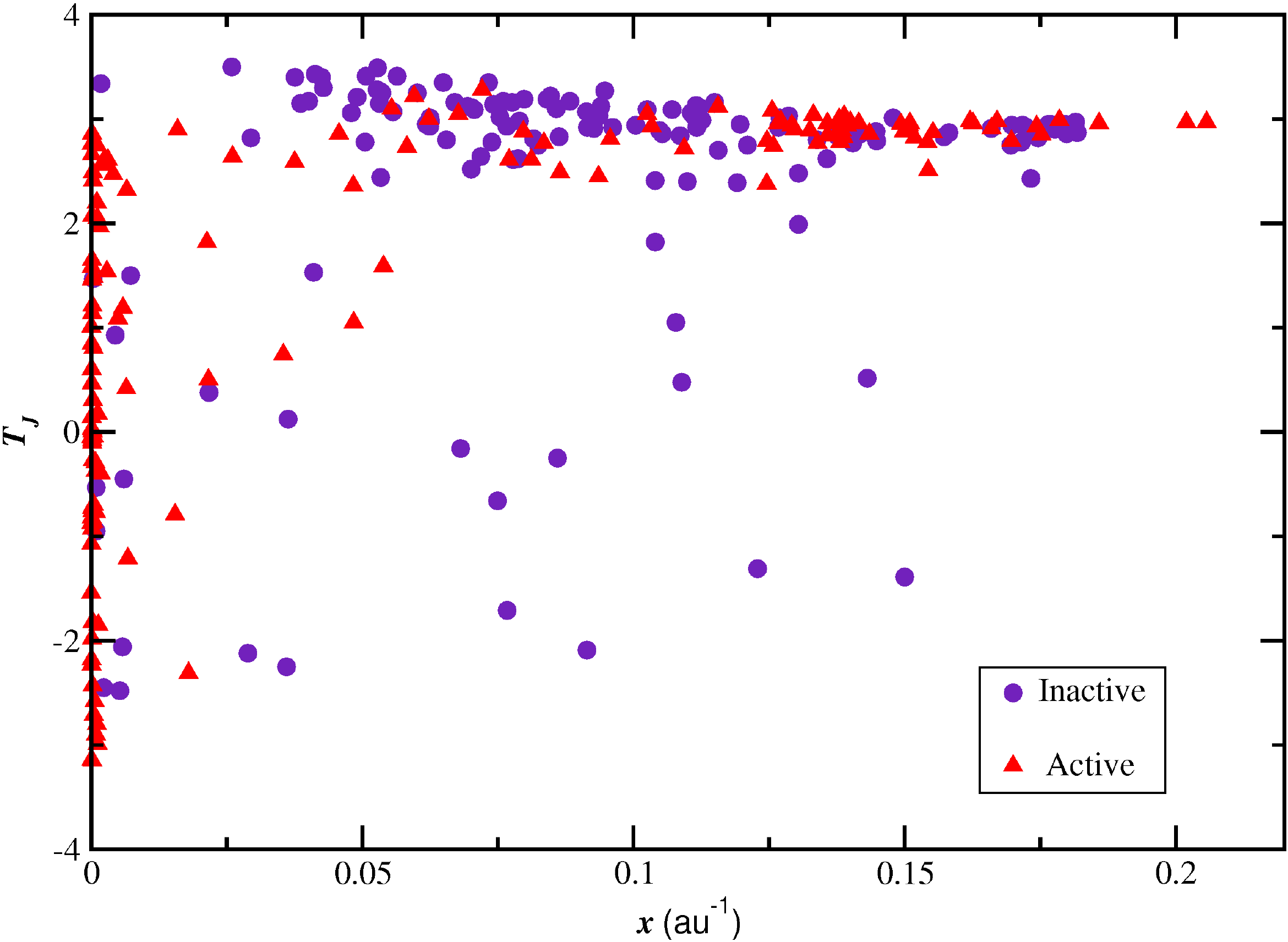}}
\caption{Active and inactive bodies with perihelion distances in the range $4 < q < 9$ au and aphelion distances $Q > 5.5$ au. Jupiter's Trojans have been removed from the sample.}
\label{x_tj}
\end{figure}

A recent drastic decrease of the Centaur's perihelion distance may also favor the display of activity, as it seems to be the case of P/2010 C1 (Scotti), a comet that shows weak activity \citep{Mazz14}, whose perihelion dropped from $q=9.50$ au to its current value of $q=5.23$ au in the last 350 yr. We found that indeed three other active Centaurs, 308P/Lagerkvist-Carsenty, P/2004 A1 (LONEOS) and P/2015 M2 (PANSTARRS) experienced drops in their perihelia from the Saturn's region to the Jupiter's region in the last few $10^2 - 10^3$ yr, whereas the $q$ of P/2005 E1 (Tubbiolo) fell from 5.93 au to 4.45 au in the last 160 yr (cf. Fig. \ref{Centauros_Memoria}). It is very suggestive that none among the inactive Centaurs was found to experience such a drastic decrease in $q$ of about a factor 1.5-2 in the recent past. We find only one inactive Centaur -2003 CC22- that shows a strong drop of its $q$ from 5.24 au to 4.16 au in the last 90 yr. It will be interesting to observe this object in the next few years (it will pass perihelion on April 25, 2023) to check whether it displays some activity as it approaches perihelion. The previous discussion strongly suggests that activity may arise as the object suddenly gets much closer to the Sun. Again, such drastic decreases in $q$ are much more likely to be found among active Centaurs than among the inactive ones because the former ones are subject on statistical terms to more frequent close encounters with the Jovian planets.

As happens many times, there are anomalous cases that inflict the general rule, as is the case of 2060 Chiron, or 95P, whose dual denomination indicates the observation of some activity. The question is, why? Chiron's properties are more in line with an object expected to be inactive: it has a long dynamical lifetime ($\sim 1.6 \times 10^6$ yr), a large size (diameter $\sim 166$ km) and a large perihelion distance that has experienced only a modest decrease in the recent past (from 8.94 au to 8.42 au in the last $\sim 1700$ yr). From occultations, \citet{Orti15} argue that Chiron may be surrounded by ring material produced by impact ejecta. The impact itself would have triggered some activity which is maintained in time through the fallback of debris on Chiron's surface and the dust and gases released from sublimating ring debris and their mutual collisions.
  
We can ask the question whether the inactive objects with Tisserand parameters $T_J < 2$ come from the Oort cloud. The capture process would require many perihelion passages during which the energy random-walks in the energy space until reaching values $x \sim 0.05-0.1$ au$^{-1}$ (semimajor axes $a \sim 10-20$ au) typical of Centaurs. We show in Fig. \ref{x_tj} a plot in the parametric plane $(x,T_J)$ for all the observed objects inactive as well as active with perihelion distances in the range $4 < q < 9$ au and aphelion distances $Q > 5.5$ au. The region with $T_J \lsim 2$ and energies $0.05 \lsim x \lsim 0.2$ au$^{-1}$ is occupied only by inactive bodies. Why are not there active bodies (``comets'')? This feature that we can describe as the {\it desert of active bodies} is rather puzzling. The only explanation that we can foresee is that fading and deactivation by exhaustion of highly volatile material, amorphous-crystalline phase transition of water ice and/or dust mantle buildup act either to disintegrate the body, or to deactivate it. The objects that can reach the ``desert'' region require a large number of passages either if they come straight from the Oort cloud, or if they evolve from the trans-neptunian region to the Oort cloud and return back to the planetary region with greater inclinations. In this regard, \citet{Fern04} found that about 50\% of the scattered disk objects end up in the Oort cloud.

The median dynamical lifetime of the Centaurs falling in the desert is $1.37 \times 10^6$ yr, as compared with $2.8 \times 10^5$ yr for the active Centaurs. It is suggestive that the amorphous/crystalline phase transition, induced by the Sun's, radiation, proceeds beneath the surface down to about 5-10 m deep where it stops \citep{Guil12}. The phase transition releases trapped molecules, like CO, contributing to the outgassing activity. \citet{Guil12} estimates that the crystallization front will stop at about $10^5$ yr at most. This is of the order of the dynamical lifetime of active Centaurs, but only one tenth of the dynamical lifetimes of Centaurs in the desert region. This may explain why the latter ones are inactive. If they were active upon arrival, their activity may have ceased long ago. Given their long dynamical time scales, the probability of finding a ``fresh'' Centaur in the desert region is very low.

The recent discoveries of two trans-neptunian objects in retrograde orbits, nicknamed Drac and Niku \citep{Glad09,Chen16}, uncover the existence of a halo population of trans-neptunian objects (TNOs) which may be a potential source of retrograde Centaurs. We cannot tell at this moment if this halo population is part of the same primordial population that gave rise to the low-inclination, disk-like TNO population, or if it originated at a different place and/or time in the solar nebula. This point is beyond the scope of this paper.

\section{Concluding remarks}

Inactive and active Centaurs have different dynamical behaviors, in particular different median dynamical lifetimes that may help to explain why some of them show notorious activity while others not. Size and perihelion distance may also contribute to explain the presence or absence of activity, as well as a recent drastic drop in perihelion distance. Probably, evolution and not intrinsic physical differences explains the display or not of activity of Centaurs in the Jupiter-Saturn region. This study shows that Centaurs with perihelia in the Jupiter-Saturn region having Tisserand parameters $T_J \lsim 2.5$ are the source of Halley-type comets, while those having $T_J \gsim 2.5$ are essentially the progenitors of JFCs.

\vspace{1cm}

{\bf Acknowledgments}

\bigskip

We acknowledge support from the Comisi\'on Sectorial de Investigaci\'on Cient\'ifica (CSIC) of the University of the Republic through the project CSIC Grupo I+D 831725 - Planetary Sciences. We thank the referees, David Jewitt and Romina di Sisto, for their helpful comments that helped to improve the presentation of the results.


\end{document}